\documentclass[preprintnumbers,article,amsmath,amssymb,floatfix,10pt,onecolumn,superscriptaddress]{revtex4-2}
\usepackage{amsmath}
\usepackage{eurosym}
\usepackage{epstopdf}
\usepackage{capt-of}
\usepackage{graphicx}
\usepackage{dcolumn}
\usepackage[left=0.8in, right=0.8in, top=1in, bottom=1in]{geometry}
\RequirePackage{graphicx}
\RequirePackage{float}
\RequirePackage{hyperref}
\RequirePackage{amsmath}
\RequirePackage{amssymb}
\RequirePackage{mathtools}
\usepackage{hyperref}
\usepackage{color}
\usepackage{comment}
\usepackage{xcolor}
\usepackage{multirow}
\usepackage{booktabs} 
\usepackage{amsmath} 
\usepackage{ragged2e}
\newcommand{\be}{\begin{equation}}
\newcommand{\ee}{\end{equation}}
\newcommand{\bea}{\begin{eqnarray}}
\newcommand{\eea}{\end{eqnarray}}
\newcommand{\eeas}{\end{eqnarray*}}
\newcommand{\beas}{\begin{eqnarray*}}

\begin{document}
\color{black}       
\title{\textbf{Effective $\Lambda$CDM model emerging from $f(Q,T)$ under a special EOS limit in symmetric cosmology with Bayesian and ANN observational constraints} }
\author{Anil Kumar Yadav}
\email{abanilyadav@yahoo.co.in}
\affiliation{Department of Physics, United College of Engineering and Research, Greater Noida - 201310, India}

\author{S. H. Shekh}
\email{da\_salim@rediff.com}
\affiliation{Department of Mathematics, S.P.M. Science and Gilani Arts, Commerce College, Ghatanji, Yavatmal, Maharashtra 445301, India}

\author{N. Myrzakulov
}\email{nmyrzakulov@gmail.com}
\affiliation{L N Gumilyov Eurasian National University, Astana 010008, Kazakhstan}

\author{Anirudh Pradhan
}\email{pradhan.anirudh@gmail.com}
\affiliation{Centre for Cosmology, Astrophysics and Space Science, GLA University, Mathura-281 406,\\ Uttar Pradesh, India}

\author{Nafis Ahmad
}\email{nafis.jmi@gmail.com}
\affiliation{Department of Physics, College of Science, King Khalid University Abha 61413, Saudi Arabia}

\author{A. M. Alshehri
}\email{amshehri@kku.edu.sa}
\affiliation{Department of Physics, College of Science, King Khalid University Abha 61413, Saudi Arabia}
\begin{abstract}
\noindent \textbf{Abstract:}  In this work, we investigate the cosmological consequences of an effective $\Lambda$CDM model emerging from the more general $f(Q,T)$ gravity theory under the special equation-of-state condition $\rho + p = 0$. Under this limit, the field equations yield the constraint $F(Q,T)H(t)=C$, implying that the function $F=f_Q$ becomes purely dependent on the non-metricity scalar $Q$, and the background evolution mimics that of the standard $\Lambda$CDM model. We derive the resulting functional forms of $f(Q)$, obtain the corresponding effective cosmological constant, and analyze the physical nature of this reduction. To test the model against observations, we perform a background-level parameter estimation for $H_{0}$ and $\Omega_{m}$, and evaluate the derived parameter $S_{8}$ (anchored to the Planck baseline $\sigma_8$) using cosmic chronometers (CC), baryon acoustic oscillations (BAO), and Pantheon+ SN Ia datasets. A dual-pipeline analysis is carried out using conventional Bayesian Markov Chain Monte Carlo (MCMC) sampling alongside a machine-learning based Artificial Neural Network (ANN) emulator deployed as a complementary consistency check. We demonstrate that the ANN approach successfully emulates the continuous parameter space with high computational efficiency, yielding results in close agreement with the standard MCMC likelihood framework. Because the background evolution strictly mimics the standard $\Lambda$CDM paradigm, the model successfully reproduces recent observational trends and recovers standard parameter constraints without dynamically resolving the baseline $H_{0}$ and $S_{8}$ tensions. Our results indicate that effective $\Lambda$CDM scenarios derived from $f(Q,T)$ gravity provide a viable and consistent with recent observational data. \\


\end{abstract}

\maketitle
\section{introduction}
Over the past few decades, an impressive body of observational evidence has confirmed that our Universe is undergoing a phase of accelerated expansion. This remarkable discovery, first indicated by high-redshift type Ia supernovae observations \cite{Riess1998,Perlmutter1999}, has been further supported by precise measurements of the cosmic microwave background (CMB) anisotropies \cite{Spergel2003,Planck2020} and large-scale structure surveys \cite{Tegmark2004,Eisenstein2005}. The origin of this acceleration remains one of the most profound questions in modern cosmology. Within the framework of general relativity (GR), it can be explained by introducing a mysterious component, dubbed \emph{dark energy}, which contributes a large negative pressure. The simplest candidate, the cosmological constant $\Lambda$, fits current observational data reasonably well \cite{Weinberg1989,Carroll2001}, but faces conceptual challenges such as the fine-tuning and cosmic coincidence problems \cite{Sahni2000,Padmanabhan2003}.

These theoretical puzzles have motivated the exploration of alternative approaches, including the modification of the gravitational sector itself. A broad class of such modifications arises from extending the Einstein-Hilbert action to more general functional forms of curvature or other geometric scalars. Notable examples include $f(R)$ gravity \cite{Nojiri2006,DeFelice2010,Sotiriou2010}, Gauss-Bonnet extensions \cite{Nojiri2005,Cognola2006}, and $f(T)$ teleparallel gravity \cite{Bengochea2009,Cai2016}. In recent years, symmetric teleparallel gravity, in which the nonmetricity scalar $Q$ plays a central role, has emerged as another viable avenue \cite{Jimenez2018,Heisenberg2019}. The corresponding $f(Q)$ theories have been shown to reproduce late-time cosmic acceleration without requiring a cosmological constant \cite{Lazkoz2019,Mandal2020} and to accommodate a wide range of cosmological behaviors.

A natural generalization of this idea is to allow the Lagrangian density to depend on both the nonmetricity scalar $Q$ and the trace $T$ of the matter energy-momentum tensor, giving rise to $f(Q,T)$ gravity \cite{Xu2019,Yousaf2020,Mandal2021}. This framework introduces an explicit coupling between matter and geometry, leading to a non-conservation of the energy-momentum tensor and consequently to extra force terms in the motion of test particles \cite{Harko2011fRT,Zubair2016}. Such couplings have been studied in related contexts like $f(R,T)$ gravity \cite{Harko2011fRT} and are known to yield rich phenomenology, from effective dark energy behavior to modifications in structure formation. The $f(Q,T)$ class thus offers a fertile ground for exploring both early- and late-universe cosmology in a unified setting.

From an observational standpoint, constraints on modified gravity theories can be obtained by confronting their predictions with cosmological data such as Hubble parameter measurements from cosmic chronometers \cite{Moresco2012,Moresco2016}, baryon acoustic oscillations (BAO) \cite{Eisenstein2005,Alam2017}, and updated type Ia supernova compilations \cite{Scolnic2018,Brout2022}. The dynamics of $f(Q,T)$ gravity can also be probed through cosmographic techniques \cite{Capozziello2008,Aviles2012}, phase-space analyses \cite{Bahamonde2017}, and model-independent reconstructions \cite{Nesseris2013}. Additionally, thermodynamic considerations in modified gravity \cite{Bamba2011,Pavon2013} can provide further theoretical guidance. Also, in recent years, symmetric teleparallel gravity has undergone rapid development, particularly through generalized extensions such as $f(Q)$ and $f(Q,T)$ theories. Several contemporary works have demonstrated that $f(Q)$ gravity can successfully reproduce late-time acceleration, provide evolving dark-energy behavior, and remain compatible with high-precision datasets. For example, reconstructions of $f(Q)$ models using non-parametric methods such as Gaussian processes have shown excellent agreement with $SNe Ia$, $BAO$, and $CC$ observations while allowing dynamical departures from $\lambda$CDM at low redshift \cite{A1, A2}. Studies using Noether symmetries and dynamical system approaches further indicate that $f(Q)$ gravity naturally admits stable accelerated attractors and scaling solutions capable of alleviating dark-energy fine-tuning problems \cite{A3, A4}. In addition, perturbation-level analyses reveal that nonmetricity-based models can modify structure growth and help reduce the S8 tension without spoiling CMB constraints \cite{A5}. Extensions toward $f(Q,T)$ gravity, which introduce explicit coupling between matter and nonmetricity, have also been actively explored. Recent investigations have examined $f(Q,T)$ cosmology in contexts such as non-linear matter interactions, anisotropic universes, compact objects, wormhole geometries, and growth index evolution, demonstrating that the T-sector can generate effective energy exchange mechanisms and nonstandard sound-speed evolution \cite{A6,A7,A8}. These developments collectively highlight that both $f(Q)$ and $f(Q,T)$ gravity provide fertile frameworks for describing early- and late-time cosmological dynamics while remaining testable against present and future survey data.

In contrast to purely geometric extensions such as $f(Q)$ gravity, the $f(Q,T)$ approach allows the matter sector to influence the effective gravitational coupling and cosmological evolution more directly. This feature is particularly interesting for addressing the cosmic coincidence problem and for generating evolving equations of state for dark energy without introducing extra scalar fields. Various functional forms of $f(Q,T)$ have been proposed, ranging from simple additive separations $f(Q,T)=f_1(Q)+f_2(T)$ \cite{Xu2019,Mandal2021} to multiplicative couplings and more elaborate parametrizations inspired by phenomenological or reconstruction-based arguments \cite{Bahamonde2021}. In the present work, we focus on the cosmological implications of $f(Q,T)$ gravity under specific matter conditions that lead to simplifications in the field equations. In particular, we consider the special cases $\rho + p = 0$ and $\rho - p = 0$, which correspond respectively to a cosmological constant-type fluid and a stiff fluid. Remarkably, for these equations of state, the background equations reduce to an effective $f(Q)$ form, with the $T$-dependence disappearing from the evolution equations. We derive the resulting constraints on $f(Q,T)$, obtain explicit functional forms in the $\rho+p=0$ case, and discuss the physical interpretation of this reduction. Importantly, we emphasize that this simplification is a consequence of the chosen matter configuration and does not imply that the underlying theory is equivalent to $f(Q)$ gravity in general. For generic fluids, the $T$-sector remains dynamically relevant, influencing both the background and perturbative behavior.\\ 
It is important to clarify the conceptual motivation behind the present construction. The aim of this work is not to introduce an alternative dynamical dark energy model in the conventional sense, but rather to investigate whether the extended framework of $f(Q,T)$ gravity can naturally accommodate standard cosmological behavior under well-defined physical conditions. In particular, by considering the limiting case $\rho + p = 0$, which corresponds to a cosmological constant-like configuration, we demonstrate that the modified field equations admit a non-trivial constraint leading to an effective reduction of the theory to a purely $Q$-dependent form. This result reveals an important and previously underexplored aspect of $f(Q,T)$ gravity: namely, that the theory is capable of reproducing $\Lambda$CDM-like expansion not by introducing an explicit cosmological constant at the level of the action, but as an emergent phenomenon arising from the interplay between geometry and matter. From this perspective, the present analysis should be viewed as a consistency test of the $f(Q,T)$ framework rather than a direct model-building exercise. The emergence of an effective cosmological constant from the underlying structure of the theory provides insight into why the $\Lambda$CDM paradigm remains remarkably successful in describing observational data, while still allowing for a deeper geometrical origin. Moreover, this approach highlights the role of matter–geometry coupling in shaping the background dynamics and establishes a bridge between modified gravity scenarios and the standard cosmological model. Therefore, the added value of the present work lies in demonstrating that $f(Q,T)$ gravity possesses the structural flexibility to recover $\Lambda$CDM behavior in a natural and internally consistent manner, without requiring the cosmological constant to be imposed a priori. Some important research on $f(Q,T)$ gravity are given in Refs. \cite{Mohanty2025,Pradhan2024,Tayde2024NPB,Tayde2024CJP,Arora2020,Arora2021}.

The paper is organized as follows: In Sec.~\ref{sec:field_eqs} we review the field equations of $f(Q,T)$ gravity. Sec.~\ref{sec:special_cases} we derived the components of field equations in the FLRW background and presents the analysis of the $\rho+p=0$ case, including the derivation of the effective $\Lambda$CDM constraints and the corresponding physical implications. Observational and theoretical aspects are discussed in Sec.~\ref{sec:observations}, and the main conclusions are summarized in Sec.~\ref{sec:conclusion}.

\section{Basic features of $f\left( Q,T\right) $ gravity}\label{sec:field_eqs}
The $f(Q,T)$ gravity framework is constructed by extending symmetric teleparallel gravity, in which gravitation is encoded not through curvature or torsion but through the nonmetricity scalar $Q$ [B2]. This formulation represents an alternative geometric description of gravity that preserves the metricity of geodesics while allowing the affine connection to remain curvature-free and torsion-free. The action for $f(Q,T)$ gravity is expressed as \cite{Xu2019}
\begin{equation}\label{e1}
S=\int \left[ L_{m} + \frac{1}{16\pi }f(Q,T)\right] \sqrt{-g}d^{4}x,
\end{equation}%
where $L_{m}$ is the   Lagrangian density of  matter, and the other symbols have their usual meanings. The definition of 
$Q$ is:
\begin{equation}\label{e2}
Q\equiv -g^{\mu \nu }\left(-L^{\delta }{}_{\alpha \delta }L^{\alpha }{}_{\mu \nu } +  L^{\delta }{}_{\alpha \mu }L^{\alpha }{}_{\nu
\delta }\right) ,
\end{equation}
and the so-called disformation tensor $L^{\delta }{}_{\alpha \gamma }$ is defined  by:
\begin{equation}\label{e3}
L^{\delta }{}_{\alpha \gamma }=\left( +\nabla _{\alpha }g_{\eta \gamma }+\nabla
_{\gamma }g_{\alpha \eta }-\nabla _{\eta
}g_{\alpha \gamma }\right)\left(-\frac{1}{2}g^{\delta \eta }\right) .
\end{equation}
We may define a non-metricity tensor as follows: form:
\begin{equation}\label{e4}
\nabla _{\gamma }g_{\mu \nu }=Q_{\gamma \mu \nu }.
\end{equation}%
Contracting the above twice wrt $g_{\mu \nu }$, we get the two non-metricity vectors:
\begin{equation}\label{e5}
Q_{\delta }=g^{\mu \nu }Q_{\delta \mu \nu },\text{ \ \ \ \ }\widetilde{Q}%
_{\delta }=g^{\mu \nu }Q_{\mu \delta \nu }. 
\end{equation}
A super-potential tensor may be defined as:
\begin{equation}\label{e6}
P_{\mu \nu }^{\delta }=+\frac{1}{4}\left(
-\widetilde{Q}^{\delta }+Q^{\delta }\right) g_{\mu \nu }-\frac{1}{4}\delta
_{(\mu }^{\delta }Q_{\nu )}-\frac{1}{2}Q_{\mu \nu }^{\delta }.
\end{equation}%
From this, the non-metricity scalar is:
\begin{equation}\label{e7}
Q=-Q_{\delta \mu \nu }P^{\delta \mu \nu }.  
\end{equation}
Varying $T_{\mu\nu}$ wrt  $%
g_{\mu \nu }$\, we get:
\begin{equation}\label{e8}
\frac{\delta (g^{\mu \nu }T_{\mu \nu })}{\delta g^{\alpha \beta }}=T_{\alpha \beta }+\theta _{\alpha \beta }.
\end{equation}
where $T_{\mu\nu}$ for matter is:
\begin{equation}\label{e9}
T_{\mu \nu }=\frac{\delta \left( \sqrt{-g}L_{m}\right) }{%
\delta g^{\mu \nu }} \frac{-2}{\sqrt{-g}},
\end{equation}%
and the variation of the matter-energy tensor with respect to the metric introduces an additional tensor $\theta_{\mu\nu}$,
\begin{equation}\label{e10}
\theta _{\mu \nu }=g^{\alpha \beta }\frac{\delta T_{\alpha \beta }}{\delta
g^{\mu \nu }}.
\end{equation}
which contributes directly to the field equations due to the explicit $T$-dependence in $f(Q,T)$. Importantly, this dependence leads to a non-conservation of the energy momentum tensor, generating extra force terms in the motion of test particles, a characteristic also observed in $f(R,T)$ theory \cite{Harko2011fRT}.\\

\noindent By varying the action (\ref{e1}) with respect to $g_{\mu\nu }$,  the field equations can be obtained  as,
\begin{equation}\label{e11}
-\frac{2}{\sqrt{-g}}\nabla _{\delta }\left( f_{Q}\sqrt{-g}P^{\delta }{}_{\mu
\nu }\right) -\frac{1}{2}fg_{\mu \nu }+f_{T}\left( T_{\mu \nu }+\theta _{\mu
\nu }\right) -f_{Q}\left( P_{\mu \delta \alpha }Q_{\nu }{}^{\delta \alpha
}-2Q^{\delta \alpha }{}_{\mu }P_{\delta \alpha \nu }\right) =8\pi T_{\mu \nu
},
\end{equation}
where $f_{T}=\frac{df\left(
Q,T\right) }{dT}$, $f_{Q}=\frac{df\left( Q,T\right) }{dQ}$, , and $\nabla _{\delta }$\ is the covariant
derivative.  Eq. (\ref{e11}) shows there is a dependence of the field equations on $\theta _{\mu \nu }$. This gives rise to a variety of models arising from $T_{\mu \nu}$. These equations encapsulate the complete dynamics of $f(Q,T)$ gravity and reveal how the geometry–matter coupling modifies both background expansion and perturbative behavior. The presence of derivatives of $f_Q$, the superpotential, and the additional tensor $\theta _{\mu \nu }$ leads to a rich phenomenology extending far beyond the minimal $f(Q)$ case. Several recent studies have explored the implications of these modifications on cosmology, structure formation, compact stars, and thermodynamic stability \cite{B4,B5,B6}.

\section{Flat FLRW model in $f\left( Q,T\right) $ theory}\label{sec:special_cases}
To explore the cosmological implications of the general field equations derived in the previous section, it is essential to implement them within a homogeneous and isotropic background. The modified dynamics encoded in Eq. (11) acquire a particularly transparent form when applied to a Friedmann–Lemaître–Robertson–Walker (FLRW) geometry. Hence, we consider the flat FLRW metric is of the form
\begin{equation}\label{e12}
ds^{2}= a^{2}(t)\left(dx^{2}+dy^{2}+dz^{2}\right)- dt^{2} ,
\end{equation}%
where $a\left( t\right) $ is the scale factor of the universe and the other symbols have their usual meanings. We will measure time in Gyr. The Hubble parameter $H$ is given by $H\equiv \dot{a}/{a}$. This implies that  $Q=6H^{2}$. We consider a perfect fluid for which: 
\begin{equation}\label{e13}
T_{\nu }^{\mu }=diag\left( -\rho ,p,p,p\right) , 
\end{equation}%
where $\rho $ is the energy density   and the pressure is $p$. So the field equations (\ref{e11}) for the metric
(\ref{e12}) yield:
\begin{equation}\label{e14}
\kappa^{2} \rho =\frac{f}{2}-6FH^{2}-\frac{2\widetilde{G}}{1+\widetilde{G}}\left( 
\overset{.}{F}H+F\overset{.}{H}\right),  
\end{equation}
\begin{equation}\label{e15}
\kappa^{2} p=-\frac{f}{2}+6FH^{2}+2\left( \overset{.}{F}H+F\overset{.}{H}\right) .
\end{equation}
where, $Q=6H^{2}$, $\left( \text{\textperiodcentered }\right) $ is $d/dt$, $\kappa^{2} \widetilde{G}\equiv f_{T}$ (here $\kappa^{2}=1$) and $F\equiv f_{Q}$  are derivatives with respect to  $T$ and $Q$,  respectively.\\
To simplify the equations, we make the substitution $	\Xi \equiv \dot{F}H + F\dot{H} = \frac{d}{dt}(F H)$ (which capture the time variation of the product $F H$.). Using this, the field equations (\ref{e14}) and (\ref{e15}) can be rewritten as
\begin{equation}\label{e16}
	\kappa^{2} \rho =\frac{f}{2}-6FH^{2}-\frac{2\widetilde{G}}{1+\widetilde{G}}\Xi ,  
\end{equation}
\begin{equation}\label{e17}
	\kappa^{2} p=-\frac{f}{2}+6FH^{2}+2\Xi.
\end{equation}
At this stage, it is worth asking whether certain simple matter conditions can make the rather general $f(Q,T)$ field equations take on a more manageable form. In some special situations, particular relations between $\rho$ and $p$ cause the combination of terms in the Friedmann equations to simplify in such a way that the dependence on $T$ effectively disappears from the background dynamics. When this happens, the theory behaves, for the purpose of the background evolution, as if it were an $f(Q)$ model, even though the underlying action is still $f(Q,T)$. This is not a universal property-it occurs only under highly symmetric conditions for the matter sector. Therefore, we choose the case of a cosmological constant-like fluid with $\rho + p = 0$. The condition $\rho+p=0$ corresponds to an equation of state parameter $\omega = -1$, mimicking a cosmological constant.\\

\noindent Adding Eqs.(\ref{e16}) and (\ref{e17}) gives
\begin{equation}\label{e18}
	\kappa^2(\rho + p) = \frac{2}{1 + \tilde{G}} \, \Xi .
\end{equation}
Imposing $\rho + p = 0$ leads directly to
\begin{equation}\label{e19}
	\Xi = 0 \quad \Rightarrow \quad \frac{d}{dt}(F H) = 0 .
\end{equation}
This integrates immediately to
\begin{equation}\label{e20}
F(Q,T) \, H(t) = C,
\end{equation}
where $C$ is a constant of integration.\\

\noindent The constraint (\ref{e20}) implies that, in order for $F(Q,T)H(t)$ to remain constant throughout the cosmological evolution, the function $F$ cannot depend explicitly on the trace $T$. Consequently, along a homogeneous and isotropic cosmological background, $F$ reduces to a function of the nonmetricity scalar $Q$ only. Hence, the explicit form of $F(Q)$ is computed as 
\begin{equation}\label{e21}
F(Q) = \frac{C\sqrt{6}}{\sqrt{Q}}.
\end{equation}
Integrating with respect to $Q$ gives
\begin{equation}\label{e22}
f(Q) = 2C\sqrt{6}\,\sqrt{Q} + B,
\end{equation}
where $B$ is an integration constant.\\

\noindent In this case, the explicit $T$-dependence of $f(Q,T)$ cancels from the background field equations due to the structure of $\rho + p = 0$, and the dynamics reduce to those of an $f(Q)$ model. But in a flat FLRW background, it is obvious that $a^{3}\sqrt{Q}\;\propto\;a^{2}\frac{da}{dt}\;\propto\;\frac{d}{dt}\left(a^{3}\right)$, so the $f(Q, T)$ part of the action Eq. (\ref{e1}) under ansatz $f(Q) = 2C\sqrt{6}\,\sqrt{Q} + B$, reduces to a total derivative and therefore does not contribute to the equations of motion. Consequently, the FLRW dynamics is governed solely by the constant $B$, and the resulting cosmological evolution is indistinguishable from the standard $\Lambda$CDM model with an effective cosmological constant $\Lambda\propto B$. Furthermore, it is worthwhile to note that earlier Paliathanasis et al. \cite{Paliathanasis/2016} have been investigated cosmological solutions of $f(T)$ gravity and obtained the similar result.\\



\noindent Using Eqs. (\ref{e20}) \& (\ref{e21}), Eq. (\ref{e22}) lead to
\begin{equation}\label{e23}
f(Q)=12C\,H + B,
\end{equation}
Inserting Eq. (\ref{e21}) and (\ref{e23}) into Eq. (\ref{e14}) for the \(\rho+p=0\), we obtain
\begin{equation}\label{e24}
\kappa^2\rho=\frac{1}{2}(12C H + B) - 6\Big(\frac{C}{H}\Big)H^2
=6C H + \frac{B}{2} - 6C H = \frac{B}{2}.
\end{equation}
Thus the energy density is constant,
\begin{equation}\label{e25}
\rho=\frac{B}{2\kappa^2}\equiv\rho_\Lambda,
\end{equation}
and plays the role of an effective cosmological constant at the background level.\\

\noindent Now, including an additional pressure-less matter component \(\rho_m\) (dust) with the usual scaling \(\rho_m(z)=\rho_{m0}(1+z)^3\), the spatially flat Friedmann equation reads
\begin{equation}\label{e26}
3H^2=\kappa^2(\rho_m+\rho_\Lambda).
\end{equation}
Define the critical density today \(\rho_{c0}\equiv 3H_0^2/\kappa^2\) and the density parameters
\(\Omega_{m}\equiv\rho_{m0}/\rho_{c0}\) and \(\Omega_\Lambda\equiv\rho_\Lambda/\rho_{c0}\), Eq. (\ref{e26}) yields
\begin{equation}\label{e28}
H(z)=H_0\sqrt{\Omega_{m}(1+z)^3+\Omega_\Lambda}.
\end{equation}
where $\Omega_{\Lambda} = 1 - \Omega_{m}$.\\

\noindent Therefore, the mapping between the integration constant \(B\) and \(\Omega_\Lambda\) are read as
\begin{equation}\label{e29}
\Omega_\Lambda=\frac{\rho_\Lambda}{\rho_{c0}}=\frac{B/(2\kappa^2)}{3H_0^2/\kappa^2}=\frac{B}{6H_0^2},
\end{equation}
or equivalently
\begin{equation}\label{e30}
B=6\,\Omega_\Lambda\,H_0^2.
\end{equation}
\noindent This gives the explicit identification of \(B\) with an effective cosmological constant. In symmetric teleparallel gravity, the nonmetricity scalar $Q$ reduces to a function of the Hubble parameter alone, which allows the gravitational modifications arising from the derivatives $f_Q$ and $f_T$ to be expressed in terms of cosmic expansion variables. This transition from the general geometric framework to the FLRW setting not only highlights how matter–geometry coupling influences the background evolution but also clarifies how specific equations of state govern the reduction to effective $\Lambda$CDM model. The following section therefore specializes the theory to a spatially flat FLRW universe, enabling an explicit evaluation of $Q$, $F=f_Q$ and $\widetilde{G}=f_T$ and leading to simplified Friedmann-like equations suitable for observational confrontation.
\begin{figure}[h]
\centering
\includegraphics[scale=0.60]{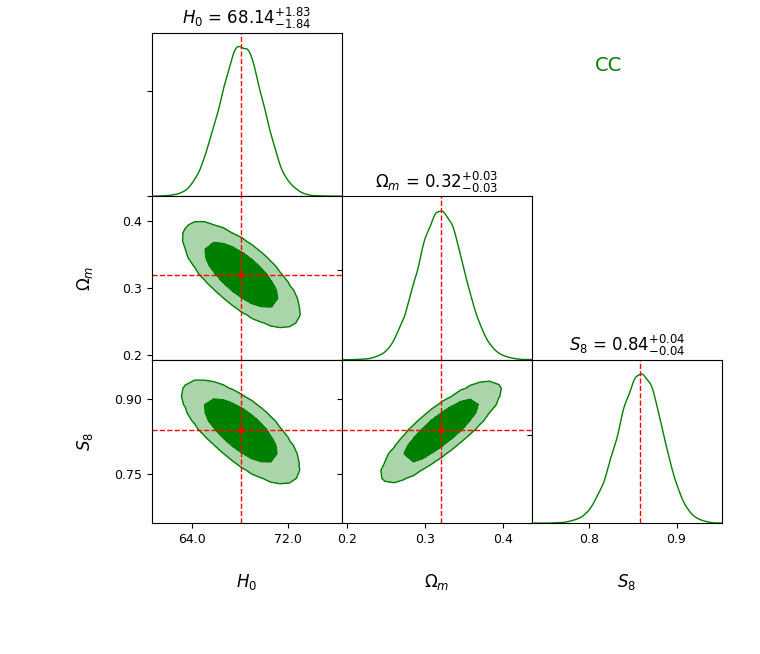}
\caption{One-dimensional marginalized probability distributions and two-dimensional ($2D$) confidence contours at $1\sigma$ and $2\sigma$ levels for the derived cosmological model, using the CC dataset analyzed with the conventional MCMC approaches.}
\end{figure}
\begin{figure}[h]
\centering
\includegraphics[scale=0.60]{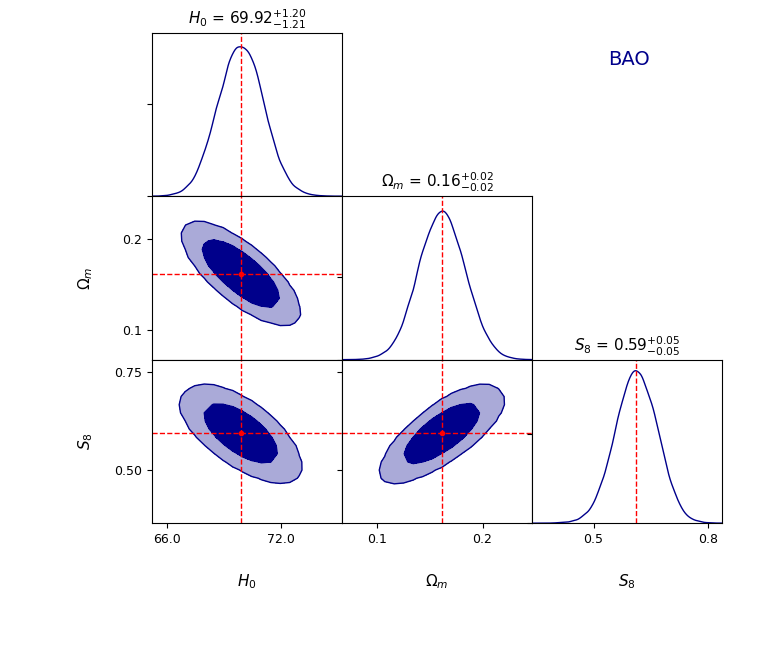}
\caption{One-dimensional marginalized probability distributions and two-dimensional ($2D$) confidence contours at $1\sigma$ and $2\sigma$ levels for the derived cosmological model, using the BAO dataset analyzed with the conventional MCMC approaches.}
\end{figure}
\begin{figure}[h]
\centering
\includegraphics[scale=0.60]{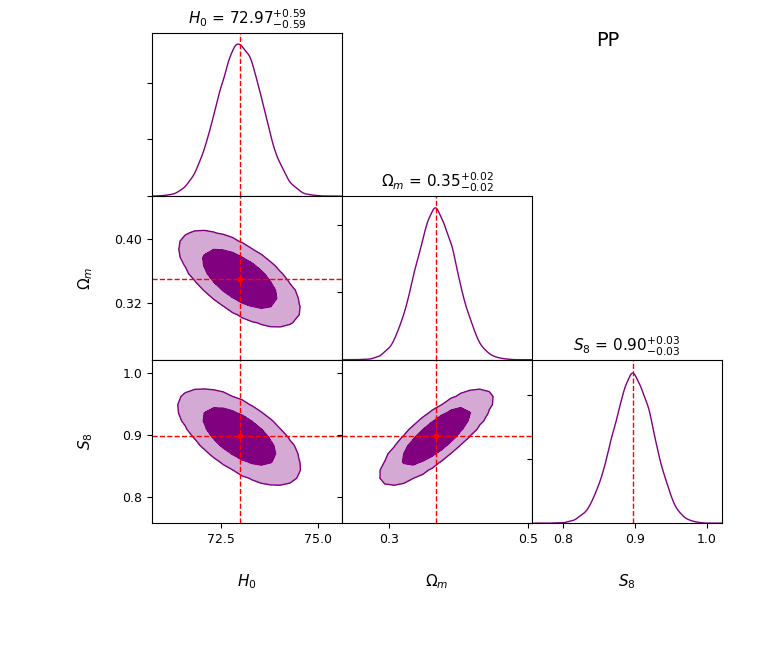}
\caption{One-dimensional marginalized probability distributions and two-dimensional ($2D$) confidence contours at $1\sigma$ and $2\sigma$ levels for the derived cosmological model, using the Pantheon+ SN Ia dataset analyzed with the conventional MCMC approaches.}
\end{figure}
\begin{figure}[h]
\centering
\includegraphics[scale=0.60]{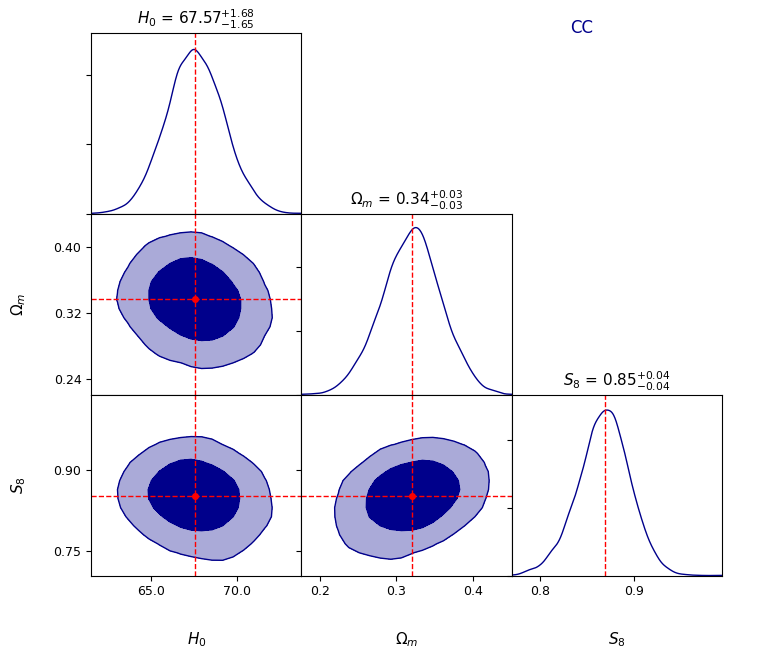}
\caption{One-dimensional marginalized probability distributions and two-dimensional ($2D$) confidence contours at $1\sigma$ and $2\sigma$ levels for the derived cosmological model, using CC dataset analyzed with advanced ANN approach.}
\end{figure}
\begin{figure}[h]
\centering
\includegraphics[scale=0.60]{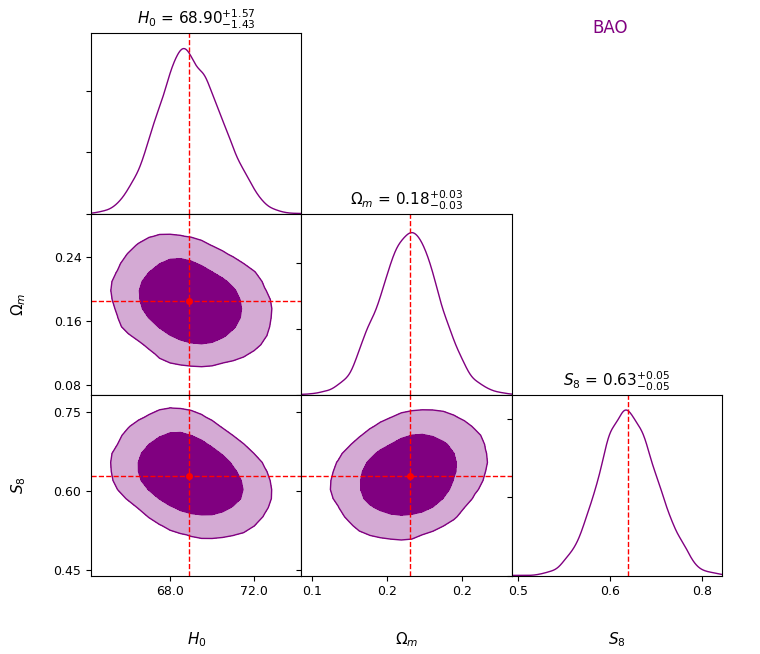}
\caption{One-dimensional marginalized probability distributions and two-dimensional ($2D$) confidence contours at $1\sigma$ and $2\sigma$ levels for the derived cosmological model, using the BAO dataset analyzed with advanced ANN approach.}
\end{figure}
\begin{figure}[h]
\centering
\includegraphics[scale=0.60]{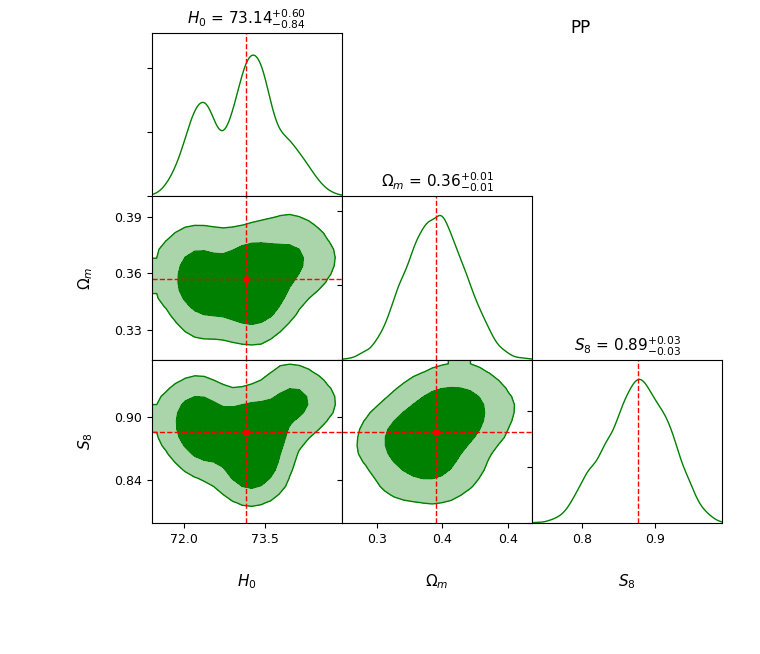}
\caption{One-dimensional marginalized probability distributions and two-dimensional ($2D$) confidence contours at $1\sigma$ and $2\sigma$ levels for the derived cosmological model, using the Pantheon+ SN Ia dataset analyzed with advanced ANN approach.}
\end{figure}
\begin{figure}[h]
\centering
\includegraphics[scale=0.50]{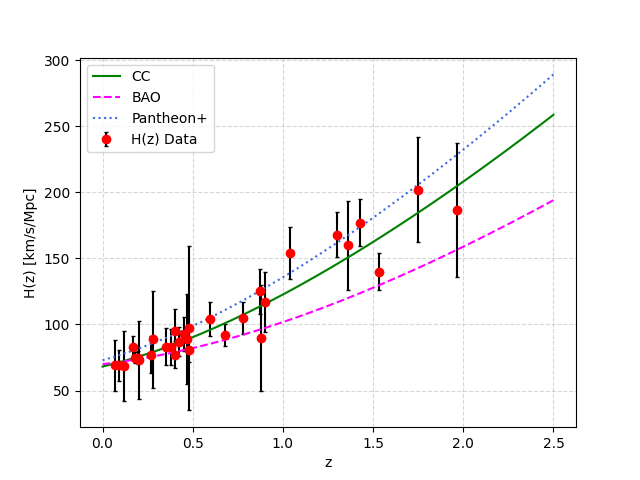}
\includegraphics[scale=0.50]{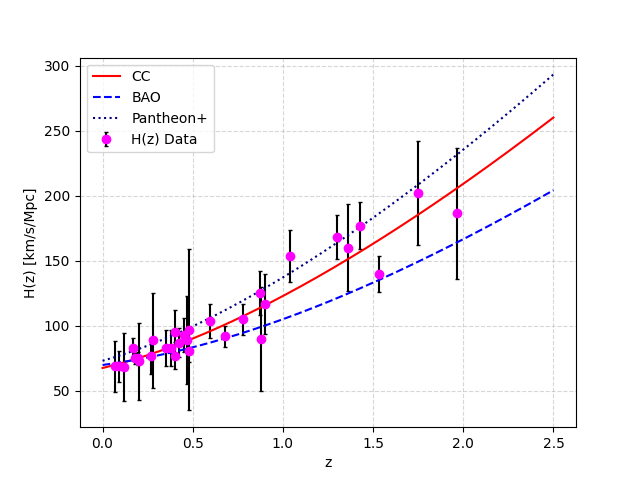}\\
\caption{The fitting of our model with $H(z)$ observed data analyzed with the conventional MCMC (left panel) \& advanced ANN  (right panel) approach.}
\end{figure}
\begin{figure}[h]
\centering
\includegraphics[scale=0.50]{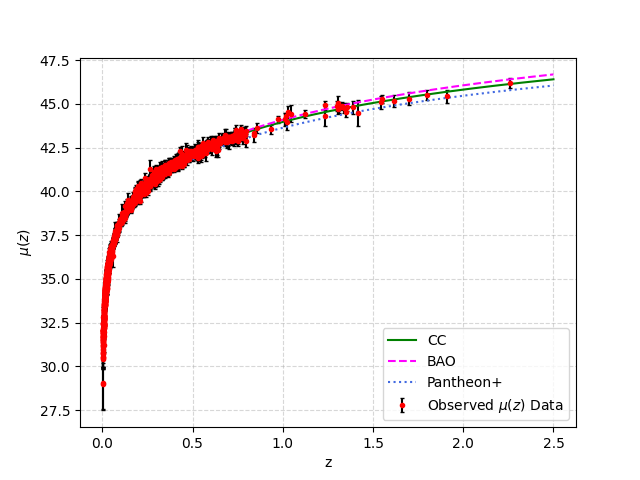}
\includegraphics[scale=0.50]{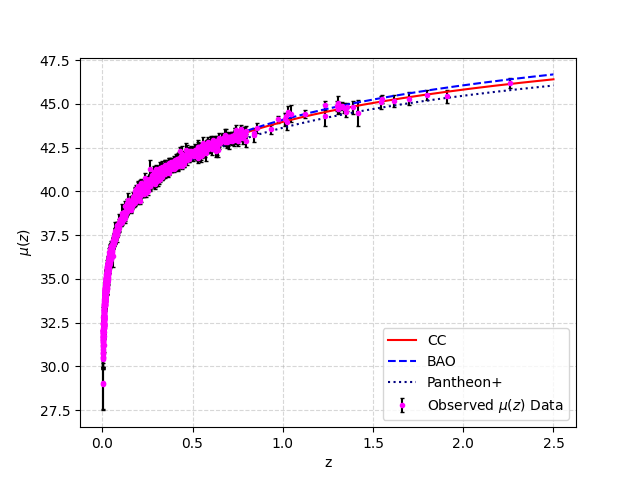}\\
\caption{The fitting of our model with Pantheon Plus compilation of SN Ia observed data analyzed with the conventional MCMC (left panel) \& advanced ANN  (right panel) approach.}
\end{figure}
\section{Confrontation with Observations}\label{sec:observations}
The functional forms obtained in the $\rho + p=0$ regimes can be tested against current cosmological datasets. Constraints from cosmic chronometer measurements of the Hubble parameter, Pantheon plus compilation of SN Ia data, and baryon acoustic oscillations are obtained. This would provide a direct connection between the theoretical framework and empirical evidence. We utilize the data sets listed below: \\

\noindent \textbf{OHD}: From the cosmic chronometric approach, we have collected $31~H(z)$ observational data points in the interval $0\leq z\leq 1.96$ \cite{Sharov/2018,Yadav/2024}.\\

\noindent \textbf{BAO data}: We used the six BAO data points given in Refs. \cite{ref49,ref50,ref51}. Moreover, the angular diameter distance is defined as $\Upsilon_{A}=\frac{\Upsilon_{\ell}}{(1+z)^{2}}$, where $\Upsilon_{\ell}$ indicates the proper angular diameter distance \cite{ref52}, and the dilation scale is described by $\Upsilon_{V}(z) = \left[ \Upsilon^{2}_{\ell}(z)*(1+z)^2*\frac{c \,  z}{H(z)} \right]^{1/3}$. \\
 
\noindent\textbf{Pantheon Plus\;(PP)}: In this work, we use the recent Pantheon+ (PP) sample that includes 1701 light curves of 1550 distinct SNe Ia in the redshift range $0.001 < z < 2.26$ \cite{Scolnic/2018,Brout/2022a}. The Pantheon+ sample is compiled from 18 different surveys, but with the SNe Ia light curves now uniformly fit to SALT2 model light curves \cite{Brout/2022b}.\\ 

\noindent The luminosity distance is computed as
\begin{equation}
D_{L}(z) = (1+z)\int_{0}^{z}\frac{c}{H(z')}\,dz',
\end{equation}
and the distance modulus is obtained from
\begin{equation}
\mu_{\mathrm{th}}(z) = 5\log_{10}\left(\frac{D_{L}(z)}{\mathrm{Mpc}}\right) + 25.
\end{equation}
\noindent The theoretical model is compared to observations through
\begin{equation}
\chi^{2} = \sum_{i=1}^{N} \frac{[E_{\mathrm{obs}}(z_i) - E_{\mathrm{th}}(z_i)]^{2}}{\sigma_{i}^{2}},
\end{equation}
where, $E_{\mathrm{obs}}$, $E_{\mathrm{th}}$ and  $\sigma_{i}$ are the observed value, theoretical value, and corresponding uncertainties in the observed value of the parameters respectively. Moreover, the model quality is quantified using $\chi^2$, Akaike Information Criterion (AIC) \cite{Burnham2004inference} and Bayesian Information Criterion (BIC) \cite{Liddle2007information}
\begin{align}
\mathrm{AIC} &= \chi^{2} + 2k,\\
\mathrm{BIC} &= \chi^{2} + k\ln N,
\end{align}
where $k$ is the number of free model parameters.

\subsection{Markov Chain Monte Carlo (MCMC) Sampling}
In order to compute the posterior probability distributions of the model parameters, we perform a Bayesian analysis using the Markov Chain Monte Carlo (MCMC) sampling technique. The posterior distribution $P(\theta|D)$ of the parameter set $\theta$ given the observational dataset $D$ is evaluated as,
\begin{equation}
P(\theta | D) = \frac{{\cal L}(D | \theta) \, \pi(\theta)}{{\cal Z}},
\label{eq:bayes}
\end{equation}
where ${\cal L}(D|\theta)$ is the likelihood function, $\pi(\theta)$ denotes the prior distribution, and ${\cal Z}$ is the normalization factor. Furthermore, we minimize $\chi^{2}$ function which corresponds to ${\cal L} \propto exp({-\chi^{2}/2})$. In this work, we use \texttt{emcee}, which implements an affine-invariant ensemble algorithm suitable for exploring correlated and degenerate parameter spaces. The resulting best fit parameter values at $1 \sigma$ uncertainties are given in Table \ref{tabl}.\\

\noindent The prior for $H_{0}$ and $\Omega_{m}$ is given as uniform within the ranges (also seen in table \ref{tab})
\begin{equation}
H_{0} \in [60, 80] \ {\rm km\,s^{-1}\,Mpc^{-1}}, \qquad
\Omega_{m} \in [0.1, 0.5].
\end{equation}

\noindent To incorporate structure growth information within the MCMC analysis, we additionally compute
\begin{equation}
S_{8} = \sigma_{8}\sqrt{\frac{\Omega_{m}}{0.3}},
\end{equation}
where $\sigma_{8}$ is the root-mean-square amplitude of matter density fluctuations on scales of $8\,h^{-1}\mathrm{Mpc}$ and we adopt here $\sigma_{8}=0.811$ from Planck results \cite{Scolnic/2018}. The parameter $S_{8}$ is evaluated at each step of the Markov chains and considered as a derived quantity which allow to check its correlation with $H_{0}$ and $\Omega_{m}$.
\begin{table}[h]
	\centering
	\caption{List of model parameters along with their prior ranges used in the analysis.}\label{tab}
	\begin{tabular}{ccc}
		\hline
		Parameter & Description & Prior Range \\
		\hline
		$H_0$ & Hubble constant (km s$^{-1}$ Mpc$^{-1}$) & $[60, 80]$ \\
		$\Omega_m$ & Matter density parameter & $[0.1, 0.5]$ \\
		\hline
	\end{tabular}
\end{table}
\subsection{Artificial Neural Network Training}
Before proceeding further, it is important to clarify the role of the Artificial Neural Network (ANN) approach adopted in this work. We emphasize that the ANN framework is not intended to replace the conventional likelihood-based inference, such as the MCMC method, which remains the standard and most rigorous approach when the likelihood function is explicitly known. In the present case, the likelihood corresponding to the observational datasets (cosmic chronometers, BAO, and Pantheon+ SNe Ia) is indeed analytically tractable, and therefore MCMC provides a reliable baseline for parameter estimation.\\
The ANN is instead employed here as a complementary tool with three specific objectives. First, it serves as a computational benchmark to explore whether machine learning-based emulators can reproduce parameter constraints with significantly reduced computational cost. Second, it provides an independent inference pipeline that allows us to test the robustness of the obtained cosmological parameters against a fundamentally different methodology. Third, it offers a means to approximate the mapping between observables and model parameters in a smooth and continuous manner, which can be useful in identifying underlying trends in parameter space.\\
From this perspective, the comparison between ANN and MCMC results should be interpreted as a consistency check rather than a competition between methods. The agreement between the two approaches indicates that the learned mapping is physically meaningful, while any differences highlight the sensitivity of parameter inference to the adopted methodology. This dual analysis strengthens the overall reliability of the results and provides additional insight into the structure of the parameter space within the $f(Q,T)$ framework.

\subsubsection{Training Strategy and Implementation Details}
To construct the ANN-based emulator, a synthetic training dataset is generated using the theoretical model predictions rather than relying solely on the limited observational data points. Specifically, we sample the parameter space $\{H_{0},\Omega_{m}\}$ within the prior ranges defined in Eq. (36), and compute the corresponding theoretical observables such as $H(z)$ and $\mu(z)$ over the relevant redshift range. This procedure ensures that the network is trained on a sufficiently dense representation of the model behavior across the parameter space. A training set is generated  as 
\begin{equation}
E_{\mathrm{train}}(z) = E_{\mathrm{th}}(z) + \mathcal{N}(0,\sigma_{\mu}).
\end{equation}
where $\mathcal{N}(0,\sigma_{\mu})$ denotes Gaussian noise using uncertainties $\sigma_{\mu}$.\\  
\noindent Therefore, the ANN input consists of full binned parameter values, while outputs correspond to physical parameters
\begin{equation}
\mathbf{y} = \{H_{0}, \Omega_{m}, S_{8}\}.
\end{equation}
\noindent Prior to training, both input and output spaces are standardized using z-score normalization
\begin{equation}
X_{\text{scaled}} = \frac{X - \langle X \rangle}{\sigma_X}.
\end{equation}
\noindent Thus, the mean network prediction produces a point estimate, while epistemic uncertainty is modeled by sampling over both ensemble members and residual variance in the scaled training space:
\begin{equation}
\theta_{\text{samples}} \sim \mathcal{N}(\hat{\theta}_{\text{ANN}}, \sigma_{\text{res}}).
\end{equation}
\noindent We draw 5000 posterior samples and transform them back to physical units
\begin{equation}
\theta_{\mathrm{phys}} = \mathrm{StandardScaler}^{-1}(\theta_{\text{samples}}).
\end{equation}
and, the resulting constrained values of parameters with $1\sigma$ errors are reported in Table 1.\\

\noindent To mimic observational uncertainties and avoid learning a purely deterministic mapping, Gaussian noise consistent with the measurement errors is added to the generated training samples according to Eq. (38). Prior to training, both input features and output parameters are standardized using z-score normalization to ensure numerical stability and efficient convergence. The network architecture consists of a feed-forward fully connected structure (Multi-Layer Perceptron). Specifically, the network is configured with an input layer corresponding to the binned observational vectors, two hidden layers containing 64 and 32 neurons respectively, and an output layer yielding the physical parameters $y=\{H_{0},\Omega_{m},S_{8}\}$. We employ the Rectified Linear Unit (ReLU) activation function for the hidden layers, while a linear activation function is applied to the output layer. The network is trained by minimizing a mean-squared error (MSE) loss function using the Adam optimizer with an initial learning rate of $\eta = 0.001$ and a batch size of 64 over 500 epochs. To further enhance robustness and handle epistemic uncertainties, we employ an ensemble approach in which 10 independent networks are trained with different weight initializations and data realizations. The final parameter estimates are obtained by averaging over the ensemble predictions, while the associated uncertainties are quantified from the spread in the ensemble outputs according to Eq. (41). This explicit architecture prevents overfitting and acts as a stable regression mapping of the background parameter space.
\begin{table}[h!]
\centering
\setlength{\tabcolsep}{5pt}
\caption{The numerical values of the parameters obtained for our model at the 68\% (1$\sigma$) confidence level, using CC, BAO and Pantheon plus compilation of SN Ia data sets analyzed with the conventional MCMC \& advanced ANN approach.}\label{tabl}
\resizebox{\textwidth}{!}{%
\begin{tabular}{lcccccccccccc}
\toprule
S.No. & Approach & Data & $H_{0}$ & $\Omega_{m}$ & $S_{8}$ & $\chi^{2}$ & AIC & BIC & $H_{0}$ Tension (Planck) & $H_{0}$ Tension (SH0ES) & $S_{8}$ Tension (Planck) & $S_{8}$ Tension (SH0ES) \\[8pt]
\midrule
1. & MCMC & CC & $68.14^{+1.83}_{-1.84}$ & $0.32^{+0.03}_{-0.03}$ & $0.84^{+0.04}_{-0.04}$ & $14.49$ & 18.49 & 21.36 & $0.39\;\sigma$ & $2.53\;\sigma$  & $0.07\;\sigma$ & $1.65\;\sigma$ \\[10pt]
2. & ANN & CC & $67.57^{+1.68}_{-1.65}$ & $0.34^{+0.03}_{-0.03}$ & $0.85^{+0.04}_{-0.04}$ & $14.59$ & 18.59 & 21.46 & $0.10\;\sigma$ & $2.95\;\sigma$  & $0.44\;\sigma$ & $1.97\;\sigma$ \\[10pt]
3. & MCMC & BAO & $69.92^{+1.20}_{-1.21}$ & $0.16^{+0.02}_{-0.02}$ & $0.59^{+0.05}_{-0.05}$ & $16.13$ & 20.13 & 20.29 & $1.92\;\sigma$ & $2.21\;\sigma$  & $4.86\;\sigma$ & $3.44\;\sigma$ \\[10pt]
4. & ANN & BAO & $68.92^{+1.57}_{-1.43}$ & $0.18^{+0.03}_{-0.03}$ & $0.63^{+0.05}_{-0.05}$ & $17.03$ & 21.03 & 21.10 & $0.95\;\sigma$ & $2.48\;\sigma$  & $4.13\;\sigma$ & $2.73\;\sigma$ \\[10pt]
5. & MCMC & Pantheon+ & $72.97^{+0.59}_{-0.59}$ & $0.35^{+0.02}_{-0.02}$ & $0.90^{+0.03}_{-0.03}$ & $745.40$ & 749.40 & 760.28 & $7.18\;\sigma$ & $0.69\;\sigma$  & $1.93\;\sigma$ & $3.87\;\sigma$ \\[10pt]
6. & ANN & Pantheon+ & $73.14^{+0.60}_{-0.84}$ & $0.36^{+0.01}_{-0.01}$ & $0.89^{+0.03}_{-0.03}$ & $747.36$ & 751.36 & 762.24 & $6.55\;\sigma$ & $0.56\;\sigma$  & $1.69\;\sigma$ & $3.76\;\sigma$ \\[10pt]
\bottomrule
\end{tabular}
}
\end{table}
\begin{figure}[h]\label{com}
\centering
\includegraphics[width=5.4cm,height=5.4cm]{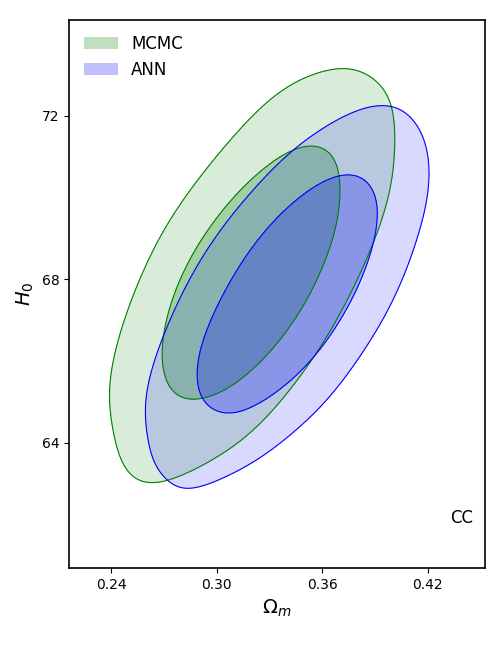}
\includegraphics[width=5.4cm,height=5.4cm]{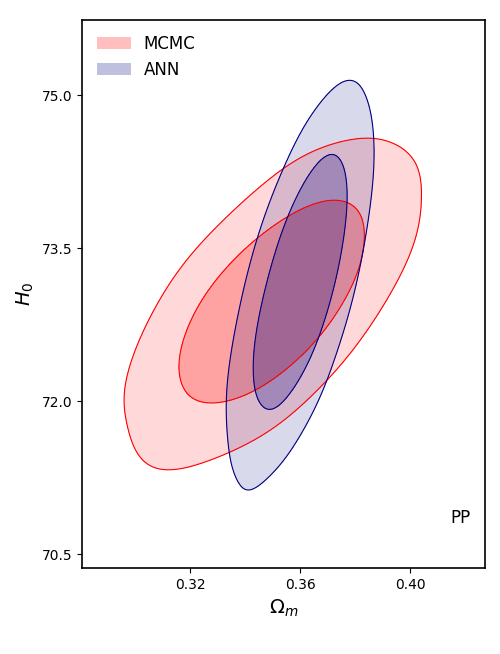}
\includegraphics[width=5.4cm,height=5.4cm]{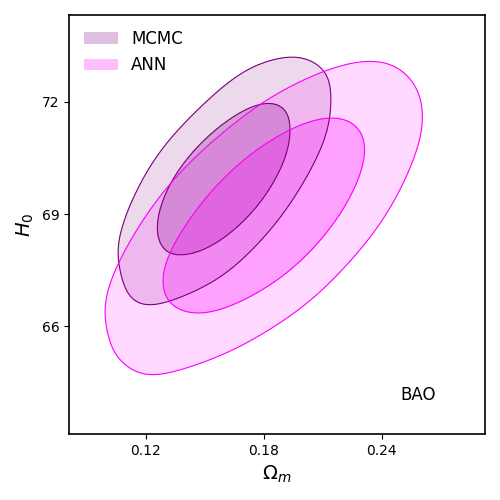}
\caption{The comparison of 2D contours in $\Omega_{m}$ - $H_{0}$ plane of our model analyzed with the conventional MCMC method \& advanced ANN approach, based on CC (left panel), Pantheon+ SN Ia (middle panel), and BAO (right panel) datasets.}
\end{figure}
\begin{figure}[h]\label{com}
\centering
\includegraphics[width=5.4cm,height=5.4cm]{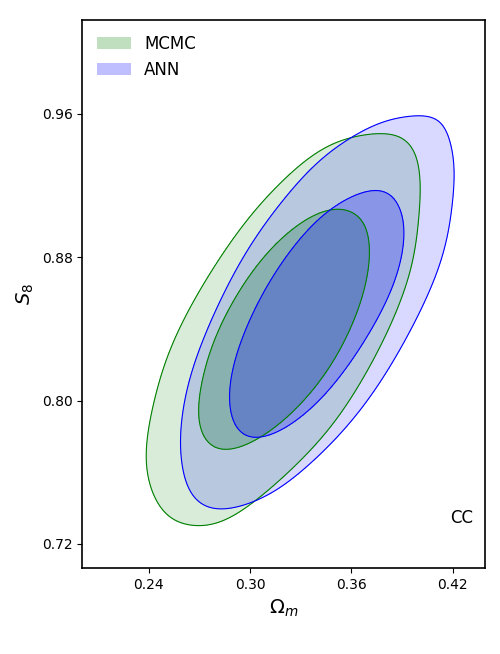}
\includegraphics[width=5.4cm,height=5.4cm]{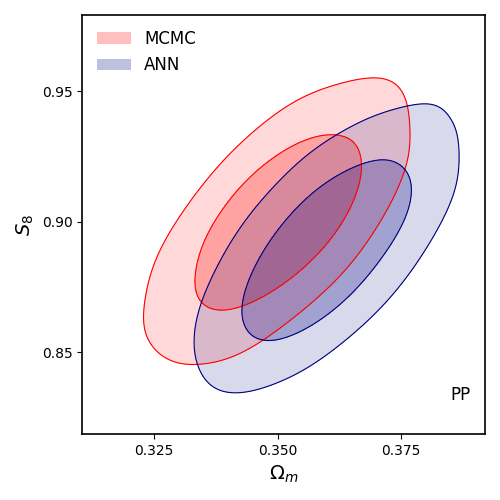}
\includegraphics[width=5.4cm,height=5.4cm]{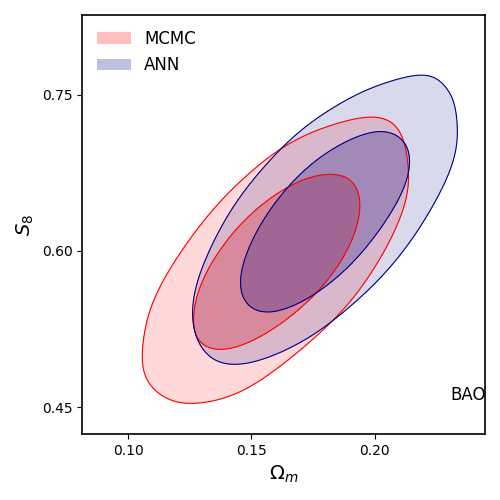}
\caption{The comparison of 2D contours in $\Omega_{m}$ - $S_{8}$ plane of our model analyzed with the conventional MCMC method \& advanced ANN approach, based on CC (left panel), Pantheon+ SN Ia (middle panel), and BAO (right panel) datasets.}
\end{figure}
\subsubsection{ On Overfitting and Interpretation of ANN Constraints}
	We acknowledge that, in general, neural network-based methods may be susceptible to overfitting, particularly when trained directly on limited observational datasets. However, in the present work, this issue is mitigated through several deliberate choices in the training procedure. First, the network is trained on a large synthetic dataset generated from the theoretical model, rather than the sparse observational data alone. This ensures that the ANN learns the global structure of the model predictions instead of memorizing individual data points. Second, the inclusion of noise in the training data acts as an effective regularization mechanism, preventing the network from over-adapting to specific realizations. Third, the use of an ensemble of networks further stabilizes the predictions and reduces variance associated with individual models.\\	
	Regarding the comparatively tighter constraints obtained from the ANN analysis, it is important to clarify that these do not arise from a fundamentally higher statistical precision than MCMC, but rather from the smooth interpolation properties of the neural network across the parameter space. The ANN effectively approximates the underlying functional relationship between observables and parameters, which can reduce sampling noise and lead to narrower posterior estimates. However, these constraints should be interpreted with caution and always in conjunction with the MCMC results, which provide the formally correct likelihood-based uncertainties.\\	
	In this sense, the ANN-derived constraints are not intended to supersede those obtained from MCMC, but to complement them by offering an alternative perspective on parameter inference. The consistency between the two methods reinforces the validity of the results, while any differences highlight the role of methodological assumptions in shaping the inferred cosmological parameters.\\
\noindent Figs. 1, 2 \& 3 show one-dimensional marginalized distributions and two-dimensional contours with $1\sigma$ and $2\sigma$ confidence levels of our model bounded with CC, Pantheon plus compilation of SN Ia data, and BAO data sets analyzed with the conventional MCMC sampling method while Figs. 4, 5 \& 6 exhibit one-dimensional marginalized distributions and two-dimensional contours with $1\sigma$ and $2\sigma$ confidence levels of our model bounded with CC, Pantheon plus compilation of SN Ia data, and BAO data sets analyzed with advanced ANN approach respectively. The $H_{0}$ is measure in km/s/Mpc. Moreover Figs. 7 \& 8 depict the fitting of our model with observed $H(z)$ data and Pantheon Plus compilation of SN Ia observed data respectively.\\ 

\noindent Fig. 9 \& 10 show the comparison of 2D contours in $\Omega_{m}$ - $H_{0}$ plane and $\Omega_{m}$ - $S_{8}$ plane of our model analyzed with the conventional MCMC method \& advanced ANN approach respectively. These plots effectively compare how two estimation methods infer cosmological parameters and shows ANN constraints are more precise, though slightly shifted relative to MCMC. The constrained values of cosmological parameter for this model using CC, BAO and Pantheon plus datasets via both the conventional MCMC technique and the advanced ANN approach are listed in Table 1. We observe that while both methods provide mutually consistent probes of $H_{0}$, $\Omega_{m}$ and $S_{8}$, the ANN method systematically gives tighter constraints, which can be evident from the reduced $1\sigma$ errors.\\

\noindent For CC data sets, we obtain $H_{0}=67.57^{+1.68}_{-1.65}$ km s$^{-1}$ Mpc$^{-1}$ from ANN approach while MCMC technique yields $H_{0} = 68.14^{+1.83}_{-1.84}$ km s$^{-1}$ Mpc$^{-1}$. These values of $H_{0}$ have $0.39\sigma$ (MCMC), $0.10\sigma$ (ANN) \& $2.53\sigma$ (MCMC), $2.95\sigma$ (ANN) with its Planck value and the local SH0ES measurement respectively. For BAO dataset, the ANN methods yields $H_{0}=68.92^{+1.57}_{-1.43}$ km s$^{-1}$ Mpc$^{-1}$ and exhibits $0.95 \sigma$ tension with Planck and $2.48 \sigma$ tension with SH0ES. The substantial $S_{8}$ tension $\sim\; 4.13\sigma$ favors the structure growth amplitudes significantly lower than KiDS and DESI estimate\cite{Heymans2021,Abbott2022}. For Pantheon+ SN Ia data, we observe that both MCMC and ANN yield a higher expansion rate $H_{0}= 72.97^{+0.59}_{-0.59}$ km s$^{-1}$ Mpc and $H_{0} = 73.14^{+0.60}_{-0.84}$ km s$^{-1}$ Mpc respectively. These values are in close agreement with the SH0ES determination \cite{Riess2022}, generating only $0.69\;\sigma$ and $0.56\;\sigma$ tension while showing strong disagreement with the Planck results \cite{Planck2018}. Moreover, we also observe that ANN method slightly reduces the $H_{0}$ tension with compare to MCMC method. The CC and BAO datasets favor Planck-like $H_{0}$, whereas Pantheon+ SN Ia data set shifts towards the SH0ES value.  Furthermore, it is worthwhile to note that The model selection criteria AIC and BIC exhibit a marginal differences between MCMC and ANN, indicating that both methods offer statistically comparable performance. As a final comment, we note that the ANN approach provides tighter constraints in maintaining parameter compatibility relative to MCMC framework.   

\subsubsection{ Advantages and Limitations of the Effective $\Lambda$CDM Reconstruction}
	In order to better position the present framework within the broader landscape of dark energy models and modified gravity theories, it is instructive to critically assess its advantages as well as its inherent limitations.\\	
	One of the primary strengths of the present approach lies in its ability to reproduce the standard $\Lambda$CDM expansion history as an emergent feature of the underlying $f(Q,T)$ gravity framework. Unlike the conventional $\Lambda$CDM model, where the cosmological constant is introduced explicitly at the level of the action, here it arises effectively from the structure of the field equations under a well-defined matter configuration. This provides a novel geometrical interpretation of dark energy, suggesting that the observed late-time acceleration may not necessarily require a fundamental constant but could instead be a manifestation of deeper matter–geometry couplings.\\	
	Another important advantage is the consistency of the model with current observational datasets. The derived Hubble expansion rate retains the same functional form as in $\Lambda$CDM, which allows for a direct and robust comparison with observational probes such as cosmic chronometers, BAO, and Type Ia supernovae. Furthermore, the incorporation of both Bayesian MCMC techniques and ANN-based inference strengthens the reliability of parameter estimation, with the latter offering improved computational efficiency and tighter constraints.\\	
	In addition, the present framework establishes a conceptual bridge between modified gravity theories and the standard cosmological paradigm. It demonstrates that extended theories such as $f(Q,T)$ gravity are not necessarily in conflict with $\Lambda$CDM, but can instead reproduce it as a limiting or effective scenario. This feature is particularly valuable in understanding the robustness of $\Lambda$CDM in light of increasingly precise cosmological observations.\\	
	However, the model also possesses certain limitations that must be acknowledged. The reduction to an effective $\Lambda$CDM behavior relies on the specific condition $\rho + p = 0$, which corresponds to a cosmological constant-like equation of state. As a result, the accelerated expansion is not dynamically generated by the theory itself but emerges under a restricted class of matter configurations. This limits the generality of the model as a fully predictive dark energy framework.\\	
	Moreover, the absence of a dynamical equation of state parameter restricts the model's ability to describe possible deviations from $\Lambda$CDM at different cosmic epochs. In contrast to other modified gravity approaches that allow for evolving dark energy behavior, the present construction effectively inherits the constant nature of $\Lambda$ once the constraint is imposed.\\	
	Finally, while the background dynamics are well captured, a complete assessment of the model would require further investigation at the perturbative level, including structure formation and growth of matter fluctuations. Such analyses are necessary to fully establish the viability of the theory beyond the homogeneous and isotropic approximation.
	
\subsubsection{ Clarification on the Interpretation of ANN Constraints}
		We would like to further clarify the interpretation of the comparatively tighter constraints obtained from the ANN analysis. These results should not be interpreted as indicating that the ANN method is statistically more rigorous or fundamentally superior to the likelihood-based MCMC approach. Instead, the apparent reduction in uncertainty arises primarily from the smoothing and interpolation properties of the neural network, which approximates the mapping between observables and model parameters in a continuous manner across the parameter space.\\		
		In contrast, the MCMC method samples the likelihood directly and therefore provides a more faithful representation of the true statistical uncertainties associated with the data. The ANN-based estimates, while consistent with MCMC results, may exhibit narrower spreads due to the learned functional approximation, rather than an intrinsic increase in statistical precision. For this reason, we emphasize that the MCMC constraints should be regarded as the primary results, while the ANN outputs serve as a complementary cross-check.\\		
		Regarding the motivation for employing ANN in the present analysis, we acknowledge that for models with analytically tractable likelihoods, such as the one considered here, traditional methods like MCMC are fully sufficient. However, the inclusion of ANN in this work is intended to explore its potential as a surrogate modeling tool, particularly in the context of future cosmological analyses where model complexity and data volume may render conventional sampling techniques computationally demanding. In this sense, the ANN framework is used here not out of necessity, but as an exploratory and comparative approach to assess its consistency with established methods.\\
		
		In summary, the present work highlights both the potential and the boundaries of the $f(Q,T)$ framework in reproducing standard cosmological behavior. While it provides a compelling geometrical interpretation of the cosmological constant as an emergent phenomenon, its applicability as a fully dynamical dark energy model remains limited to specific physical conditions. Future studies aimed at relaxing these constraints and exploring more general scenarios may offer deeper insights into the role of non-metricity and matter coupling in cosmic evolution.

\section{Conclusion}\label{sec:conclusion}
In this paper, we investigated effective $\Lambda$CDM model emerging from $f(Q,T)$ under condition $\rho + p = 0$ which yields a constraint relation $F(Q,T)H(t)=C$, and the function $F$ reduces to a purely $Q$-dependent form as$F(Q)=C\sqrt{6/Q}$ along the homogeneous and isotropic cosmological background. This reduction from $f(Q,T)$ gravity to an effective $\Lambda$ formulation indicates that the trace of the energy momentum tensor evolves toward a constant value, thereby mimicking the behavior of a cosmological constant. We also emply both the Bayesian statistical inference and Artificial Neural Network (ANN) emulator to constraint the model parameters with CC, BAO and Pantheon+ SN Ia data sets. This ANN approach significantly reduces computational time and enhances the robustness of parameter estimation procedures. In Ref. \cite{Escamilla-Rivera2020}, a machine-learning approach have been discussed to improve performance in modified theory of gravity. The main findings of this research are as follows:

\begin{itemize}
\item[i)] We derived an effective $\Lambda$CDM model from the general $f(Q,T)$ gravity by imposing the condition $\rho + p = 0$ which yields $F(Q,T)H(t)=\mathrm{constant}$.

\item[ii)] For the CC dataset, our model exhibits $H_{0}$ tensions of $0.39\sigma$ (MCMC) and $0.10\sigma$ (ANN) with Planck, and $2.53\sigma$ (MCMC) and $2.95\sigma$ (ANN) with SH0ES and shows an insignificant $S_{8}$ tension with Planck which imply that CC data set alone does not strongly probe late-time structure formation of the Universe. 
    
\item[iii)] For the BAO dataset, the ANN method yields a $0.95\sigma$ $H_{0}$ tension with Planck and a $2.48\sigma$ tension with SH0ES and a significant $S_{8}$ tension of $\sim 4.13\sigma$, favoring lower structure growth amplitudes relative to KiDS and DESI \cite{Heymans2021,Abbott2022}.
    
\item[iv)] For the Pantheon+ SN Ia dataset, we obtain $H_{0}=72.97^{+0.59}_{-0.59}$ km s$^{-1}$ Mpc$^{-1}$ (MCMC) \& $H_{0}=73.14^{+0.60}_{-0.84}$ km s$^{-1}$ Mpc$^{-1}$ (ANN) which are consistent with the SH0ES result \cite{Riess2022}. 
    
\item[v)] The CC and BAO datasets favor Planck-like values of $H_{0}$, whereas Pantheon+ SN Ia data prefers SH0ES-like values, highlighting the persistent $H_{0}$ tension.
    
\item[vi)] The ANN approach  slightly alleviates the $H_{0}$ tension compared to MCMC and provides tighter parameter constraints.
    
\item[vii)] The AIC and BIC show only marginal differences between ANN and MCMC approaches, indicating statistically comparable performance.

\end{itemize} 
 
Importantly, to clarify the treatment of the parameter $S_8$ in the present analysis. The quantity $S_8 = \sigma_8 \sqrt{\Omega_m/0.3}$ is intrinsically related to the growth of matter perturbations and, in a fully consistent modified gravity framework, should ideally be derived from the evolution of linear density fluctuations. In this work, however, our analysis is primarily restricted to the background cosmological dynamics, and a complete perturbative treatment of $f(Q,T)$ gravity has not been performed. Consequently, the value of $\sigma_8$ is not derived within the model itself, but is instead adopted from Planck observations as a reference value. This approach allows us to estimate $S_8$ as an indicative parameter for comparison with observational constraints, rather than as a fully model-dependent prediction. We note that, under the condition $\rho + p = 0$, the $f(Q,T)$ framework effectively reduces to a $\Lambda$CDM-like expansion at the background level. In this limit, it is reasonable to expect that the growth of structures does not deviate significantly from the standard scenario, at least to leading order. This provides a partial justification for using the Planck value of $\sigma_8$ as an approximation. Nevertheless, we emphasize that a rigorous determination of $S_8$ within the $f(Q,T)$ framework would require a detailed perturbation analysis, including the study of growth equations and effective gravitational coupling. Such an investigation lies beyond the scope of the present work and will be considered in future studies.\\

It is important to clarify the scope and implications of the present model in the context of existing cosmological tensions. Since the derived background evolution coincides with that of the standard $\Lambda$CDM model, one should not expect a direct resolution of the $H_0$ or $S_8$ tensions within the framework considered here. These tensions typically arise from discrepancies between different observational probes and often require departures from the standard expansion history or modifications at the perturbative level. The primary objective of this work is therefore not to propose a competing model that resolves such tensions, but rather to demonstrate that the $f(Q,T)$ gravity framework is capable of reproducing the $\Lambda$CDM behavior as an effective and emergent scenario under specific physical conditions. In this sense, the significance of the model lies in providing a deeper theoretical understanding of the robustness of the $\Lambda$CDM paradigm, despite its well-known conceptual challenges. Furthermore, the use of standard observational datasets and conventional statistical techniques, such as MCMC analysis, is intentional. This allows for a transparent and direct comparison with existing results in the literature and ensures that the consistency of the model with observations is assessed within a well-established framework. The inclusion of the ANN-based approach, on the other hand, serves as an additional exploratory tool to examine the stability of parameter estimation under an alternative inference methodology. Thus, rather than aiming to replace $\Lambda$CDM or to directly resolve current observational tensions, the present work contributes by establishing a link between modified gravity theories and the standard cosmological model, highlighting how effective $\Lambda$-like behavior can arise from underlying geometrical structures.\\

Finally, our findings indicate that the effective $\Lambda$CDM scenario derived from $f(Q,T)$ represents a viable extension of non-metricity gravity capable of addressing late-time cosmic acceleration while remaining fully aligned with current observational constraints. The ANN framework demonstrates improved efficiency in narrowing confidence intervals while maintaining dataset consistency. Future high-precision cosmological surveys (e.g., Euclid, LSST) will be crucial for further testing such deviations and for distinguishing effective non-metricity dark energy contributions from from $\Lambda$CDM behavior.
 
\section*{Declaration of competing interest}
The authors declare that they have no known competing financial interests or personal relationships that could have appeared to influence the work reported in this paper.

\section*{Data availability}
No data was used for the research described in the article.

\section*{Acknowledgment}
The authors sincerely thank the anonymous referee for the careful reading of the manuscript and for the constructive comments and valuable suggestions. These insightful remarks have significantly contributed to improving the clarity, quality, and overall presentation of the paper. The authors (AKY, NA \& AMA) extend their appreciation to the Deanship of Research and Graduate Studies at King Khalid University, Saudi Arabia for funding this work through large group Research Project under grant number RGP. 2/310/47. The authors express their gratitude to Prof. A. Paliathanasis for the constructive comments and suggestions that contributed to improving this work. 

\end{document}